%% file: lcwsSLACHiggs.tex
\documentclass[slac_one]{revtex4}
\usepackage{graphicx}
\usepackage{epsfig}
\usepackage{fancyhdr}
\usepackage{amsfonts}
\usepackage{amsmath}
\usepackage{amssymb}
\usepackage{tabularx}
\usepackage{feynarts}
\usepackage[dvips]{color}
\usepackage{colordvi,psfrag}
\def\alt{\alpha_t}

\input paperdef

\begin{document}

% CERN-PH-TH/2005-113
% MPP-2005-67
% DCPT/05/DD
% IPPP/05/II

\title{{\small{2005 International Linear Collider Workshop - Stanford,
U.S.A.}}\\ %% Please keep this conference title here
\vspace{12pt}
The MSSM Higgs Sector at \boldmath{\order{\alb\als}} and Beyond} %% 

\author{S.~Heinemeyer}
\affiliation{CERN TH Division, Department of Physics,
CH-1211 Geneva 23, Switzerland}
\author{W.~Hollik, H.~Rzehak}
\affiliation{Max-Planck-Institut f\"ur Physik (Werner-Heisenberg-Institut),
F\"ohringer Ring 6, D--80805 Munich, Germany}
\author{G.~Weiglein}
\affiliation{Institute for Particle Physics Phenomenology, University
of Durham, Durham DH1~3LE, UK}

\begin{abstract}
We evaluate \order{\alb\als} corrections in the MSSM Higgs boson
sector, including an analysis of the renormalization in the 
bottom/scalar bottom sector.
For $\mu < 0$ the genuine two-loop corrections to the mass of the
lightest Higgs boson mass can amount up to 3~GeV.
Different renormalization schemes are applied and numerically compared. 
Likewise the residual dependence on the renormalization scale
is investigated. This allows to determine the remaining theoretical
uncertainties from unknown higher-order corrections at
\order{\alb\als^2} for different regions of the MSSM parameter space.
\end{abstract}

\maketitle

\thispagestyle{fancy}

%%%%%%%%%%%%%%%%%%%%%%%%%%%%%%%%%%%%%%%%%%%%%%%%%%%%%%%%%%%%%%%%%%%%%%%%%%%%%%%
%%%%%%%%%%%%%%%%%%%%%%%%%%%%%%%%%%%%%%%%%%%%%%%%%%%%%%%%%%%%%%%%%%%%%%%%%%%%%%%

\section{INTRODUCTION} 

A crucial prediction of the Minimal Supersymmetric Standard Model
(MSSM)~\cite{susy} is the existence of at least one light Higgs
boson. Direct searches at LEP have already ruled out a
considerable fraction of the MSSM parameter
space~\cite{LEPHiggsSM,LEPHiggsMSSM}, and the forthcoming
high-energy experiments at the Tevatron, the LHC, and the International
Linear Collider (ILC) will either
discover a light Higgs boson or rule out Supersymmetry (SUSY) 
as a viable theory
for physics at the weak scale. Furthermore, if one or more Higgs
bosons are discovered, bounds on their masses and couplings will be 
set at the LHC~\cite{LHCHiggs,HcoupLHCSM}. 
Eventually the masses and couplings will be determined
with high accuracy at the ILC~\cite{tesla,orangebook,acfarep}. 
Thus, a
precise knowledge of the dependence of masses and mixing angles in
the MSSM Higgs sector on the relevant supersymmetric parameters is of
utmost importance to reliably compare the predictions of the MSSM with
the (present and future) experimental results.
Even nowadays a precise prediction of the lightest MSSM Higgs boson
mass is important to set solid bounds on the MSSM parameter
space~\cite{tbexcl,PomssmRep}. 

The status of the available results for the higher-order contributions to
the neutral $\cp$-even MSSM Higgs boson masses has been summarized in
\citeres{mhiggsAEC,habilSH,mhiggsAWB}. 
In particular two-loop contributions to the leading one-loop
corrections involving the top and bottom Yukawa couplings (with 
$\alt \equiv h_t^2 / (4 \pi)$, $\alb \equiv h_b^2 / (4\pi)$,
$h_{t,b}$ being the superpotential top or bottom coupling) have been
evaluated. 
Corrections from the bottom/sbottom sector can also give large effects,
in particular for large values of $\tb$, the ratio of the two vacuum
expectation values, $\tb = v_2/v_1$, and large values of $\mu$, 
the supersymmetric Higgs mass parameter.
The leading strong corrections at \order{\alb\als} have been derived
in \citere{mhiggsEP4} (in the limit $\tb \to \infty$) and in
\citere{mhiggsFDalbals} (for arbitrary $\tb$).
In the (s)bottom
corrections the all-order resummation of the $\tb$-enhanced terms,
\order{\alb(\als\tb)^n}, is also performed \cite{deltamb1,deltamb}.
Furthermore a full effective potential two-loop calculation
exists~\cite{effpotfull}, however, no public code is available.

The potentially large size of corrections from the $b/\Sbot$ sector
makes it desirable to investigate the
corresponding two-loop corrections, including the applied 
renormalization. An
inconvenient scheme can give rise to artificially large corrections,
whereas a convenient scheme absorbs the dominant contributions into the
one-loop result, and higher-order corrections remain small. The
comparison of different schemes (without artificially enhanced
corrections)~and the renormalization scheme dependence
give an indication of the possible size of missing 
higher-order terms of~\order{\alb\als^2}.

%%%%%%%%%%%%%%%%%%%%%%%%%%%%%%%%%%%%%%%%%%%%%%%%%%%%%%%%%%%%%%%%%%%%%%%%%%%%%%%
%%%%%%%%%%%%%%%%%%%%%%%%%%%%%%%%%%%%%%%%%%%%%%%%%%%%%%%%%%%%%%%%%%%%%%%%%%%%%%%

\section{THE MSSM HIGGS SECTOR AND RENORMALIZATION}
\label{sec:higgssector}

The Higgs sector of the MSSM~\cite{hhg} comprises two
neutral $\cp$-even Higgs bosons, $h$ and $H$ ($\mh < \mH$), the
$\cp$-odd $A$~boson (throughout this paper we assume that $\cp$ is
conserved) and two charged Higgs bosons, $H^\pm$.

In the Feynman-diagrammatic (FD) approach, the higher-order corrected 
Higgs boson masses, $\Mh$ and $\MH$, are derived as the poles of the
$h,H$-propagator matrix,  
i.e.\ by solving the equation
\begin{equation}
\left[p^2 - \mhtree^2 + \hSi_{hh}(p^2) \right]
\left[p^2 - \mHtree^2 + \hSi_{HH}(p^2) \right] -
\left[\hSi_{hH}(p^2)\right]^2 = 0\,.
\label{eq:proppole}
\end{equation}
The renormalized self-energies, $\hSi_{s}$, 
can be expanded according to 
the one-, two-, \ldots loop-order contributions,
\begin{equation}
\label{renSE}
\hSi_{s}(p^2) = \hSi_{s}^{(1)}(p^2) 
                   + \hSi_{s}^{(2)}(p^2) + \cdots ~, s = hh, hH, HH~.
\end{equation}
The leading \twol\ corrections from the $b/\Sbot$~sector are the
\order{\als} corrections to the dominant \onel\ contributions of
\order{\alb}. They 
are obtained for zero external momentum and
neglecting the gauge couplings, see \citere{mhiggsFDalbals} for
further details.
This approach is analogous to the way the
leading one- and \twol\ contributions in the top/stop sector have been
obtained (see e.g.~\citere{mhiggslong}).
Details about the renormalization in the MSSM Higgs sector can be
found in \citere{mhiggsFDalbals}.

%%%%%%%%%%%%%%%%%%%%%%%%% F I G U R E %%%%%%%%%%%%%%%%%%%%%%%%%%%%%%%%%%%%%%%%%
\vspace{-4em}
\begin{figure}[htb!]
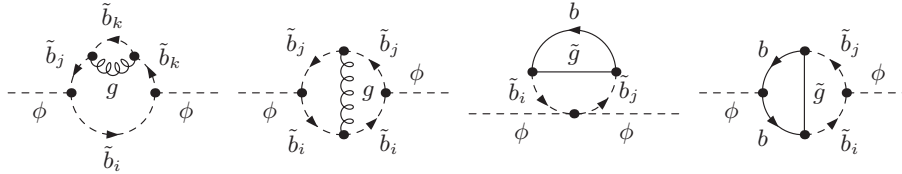

\unitlength=1bp%

\begin{feynartspicture}(453,113)(4,1.3)

\FADiagram{}
\FAProp(0.,10.)(6.,10.)(0.,){/ScalarDash}{0}
\FALabel(3.,9.18)[t]{$\phi$}
\FAProp(20.,10.)(14.,10.)(0.,){/ScalarDash}{0}
\FALabel(17.,9.18)[t]{$\phi$}
\FAProp(6.,10.)(14.,10.)(1.,){/ScalarDash}{1}
\FALabel(10.,4.93)[t]{$\tilde b_i$}
\FAProp(8.,13.5)(6.,10.)(0.315846,){/ScalarDash}{1}
\FALabel(5.58149,12.3549)[br]{$\tilde b_j$}
\FAProp(12.,13.5)(8.,13.5)(0.8,){/ScalarDash}{1}
\FALabel(10.,16.17)[b]{$\tilde b_k$}
\FAProp(12.,13.5)(8.,13.5)(-0.8,){/Cycles}{0}
\FALabel(10.,10.83)[t]{$g$}
\FAProp(12.,13.5)(14.,10.)(-0.310119,){/ScalarDash}{-1}
\FALabel(14.4085,12.3491)[bl]{$\tilde b_k$}
\FAVert(12.,13.5){0}
\FAVert(8.,13.5){0}
\FAVert(6.,10.){0}
\FAVert(14.,10.){0}

\FADiagram{}
\FAProp(0.,10.)(6.,10.)(0.,){/ScalarDash}{0}
\FALabel(3.,9.18)[t]{$\phi$}
\FAProp(20.,10.)(14.,10.)(0.,){/ScalarDash}{0}
\FALabel(17.,10.82)[b]{$\phi$}
\FAProp(10.,6.)(6.,10.)(-0.434885,){/ScalarDash}{-1}
\FALabel(6.51421,6.51421)[tr]{$\tilde b_i$}
\FAProp(10.,6.)(14.,10.)(0.412689,){/ScalarDash}{1}
\FALabel(13.4414,6.55861)[tl]{$\tilde b_i$}
\FAProp(10.,14.)(10.,6.)(0.,){/Cycles}{0}
\FALabel(11.77,10.)[l]{$g$}
\FAProp(10.,14.)(6.,10.)(0.425735,){/ScalarDash}{1}
\FALabel(6.53252,13.4675)[br]{$\tilde b_j$}
\FAProp(10.,14.)(14.,10.)(-0.412689,){/ScalarDash}{-1}
\FALabel(13.4414,13.4414)[bl]{$\tilde b_j$}
\FAVert(10.,14.){0}
\FAVert(10.,6.){0}
\FAVert(6.,10.){0}
\FAVert(14.,10.){0}

\FADiagram{}
\FAProp(0.,8.)(10.,8.)(0.,){/ScalarDash}{0}
\FALabel(5.,7.18)[t]{$\phi$}
\FAProp(20.,8.)(10.,8.)(0.,){/ScalarDash}{0}
\FALabel(15.,7.18)[t]{$\phi$}
\FAProp(6.,12.)(14.,12.)(-1.,){/Straight}{-1}
\FALabel(10.,17.07)[b]{$b$}
\FAProp(6.,12.)(14.,12.)(0.,){/Straight}{0}
\FALabel(10.,12.72)[b]{$\tilde g$}
\FAProp(6.,12.)(10.,8.)(0.431748,){/ScalarDash}{1}
\FALabel(3.75252,9.05252)[bl]{$\tilde b_i$}
\FAProp(14.,12.)(10.,8.)(-0.450694,){/ScalarDash}{-1}
\FALabel(16.2854, 9.01463)[br]{$\tilde b_j$}
\FAVert(6.,12.){0}
\FAVert(14.,12.){0}
\FAVert(10.,8.){0}

\FADiagram{}
\FAProp(0.,10.)(6.,10.)(0.,){/ScalarDash}{0}
\FALabel(3.,9.18)[t]{$\phi$}
\FAProp(20.,10.)(14.,10.)(0.,){/ScalarDash}{0}
\FALabel(17.,10.82)[b]{$\phi$}
\FAProp(10.,6.)(6.,10.)(-0.434885,){/Straight}{-1}
\FALabel(6.51421,6.51421)[tr]{$b$}
\FAProp(10.,6.)(14.,10.)(0.412689,){/ScalarDash}{1}
\FALabel(13.4414,6.55861)[tl]{$\tilde b_i$}
\FAProp(10.,14.)(10.,6.)(0.,){/Straight}{0}
\FALabel(10.82,10.)[l]{$\tilde g$}
\FAProp(10.,14.)(6.,10.)(0.425735,){/Straight}{1}
\FALabel(6.53252,13.4675)[br]{$b$}
\FAProp(10.,14.)(14.,10.)(-0.412689,){/ScalarDash}{-1}
\FALabel(13.4414,13.4414)[bl]{$\tilde b_j$}
\FAVert(10.,14.){0}
\FAVert(10.,6.){0}
\FAVert(6.,10.){0}
\FAVert(14.,10.){0}

\end{feynartspicture}
\vspace{-1em}
\caption{Some generic \twol\ diagrams for the Higgs-boson self-energies
($\phi = h, H, A$;
$\;i,j,k,l = 1,2$). Similar diagrams arise for the tadpole contributions.}
\label{fig:FD2L}
%\end{center}
\end{figure}
%%%%%%%%%%%%%%%%%%%%%%%% F I G U R E %%%%%%%%%%%%%%%%%%%%%%%%%%%%%%%%%%%%%%%%%

%%%%%%%%%%%%%%%%%%%%%%%% F I G U R E %%%%%%%%%%%%%%%%%%%%%%%%%%%%%%%%%%%%%%%%%
\vspace{-4em}
\begin{figure}[htb]
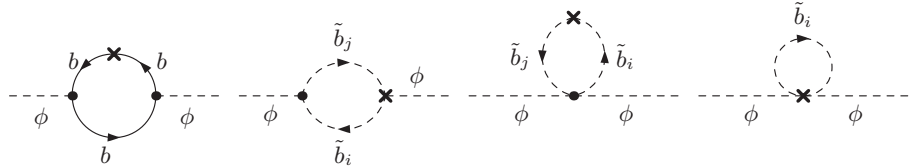

\begin{feynartspicture}(453,113)(4,1.3)

\FADiagram{}
\FAProp(0.,10.)(6.,10.)(0.,){/ScalarDash}{0}
\FALabel(3.,8.9)[t]{$\phi$}
\FAProp(20.,10.)(14.,10.)(0.,){/ScalarDash}{0}
\FALabel(17.,8.9)[t]{$\phi$}
\FAProp(10.,14.)(6.,10.)(0.431295,){/Straight}{1}
\FALabel(6.60271,12.5973)[br]{$b$}
\FAProp(10.,14.)(14.,10.)(-0.431295,){/Straight}{-1}
\FALabel(14.0675,12.5675)[bl]{$b$}
\FAProp(6.,10.)(14.,10.)(1.,){/Straight}{1}
\FALabel(9.2,5.18)[t]{$b$}
\FAVert(6.,10.){0}
\FAVert(14.,10.){0}
\FAVert(10.,14.){1}

\FADiagram{}
\FAProp(0.,10.)(6.,10.)(0.,){/ScalarDash}{0}
\FALabel(3.,9.18)[t]{$\phi$}
\FAProp(20.,10.)(14.,10.)(0.,){/ScalarDash}{0}
\FALabel(17.,10.82)[b]{$\phi$}
\FAProp(6.,10.)(14.,10.)(0.8,){/ScalarDash}{-1}
\FALabel(10.,5.73)[t]{$\tilde b_i$}
\FAProp(6.,10.)(14.,10.)(-0.8,){/ScalarDash}{1}
\FALabel(10.,14.27)[b]{$\tilde b_j$}
\FAVert(6.,10.){0}
\FAVert(14.,10.){1}

\FADiagram{}
\FAProp(0.,10.)(10.,10.)(0.,){/ScalarDash}{0}
\FALabel(5.,9.18)[t]{$\phi$}
\FAProp(20.,10.)(10.,10.)(0.,){/ScalarDash}{0}
\FALabel(15.,9.18)[t]{$\phi$}
\FAProp(10.,17.5)(10.,10.)(-0.8,){/ScalarDash}{-1}
\FALabel(14.07,13.75)[l]{$\tilde b_i$}
\FAProp(10.,17.5)(10.,10.)(0.8,){/ScalarDash}{1}
\FALabel(5.93,13.75)[r]{$\tilde b_j$}
\FAVert(10.,10.){0}
\FAVert(10.,17.5){1}

\FADiagram{}
\FAProp(0.,10.)(10.,10.)(0.,){/ScalarDash}{0}
\FALabel(5.,9.18)[t]{$\phi$}
\FAProp(20.,10.)(10.,10.)(0.,){/ScalarDash}{0}
\FALabel(15.,9.18)[t]{$\phi$}
\FAProp(10.,10.)(10.,10.)(10.,15.5){/ScalarDash}{-1}
\FALabel(10.,16.57)[b]{$\tilde b_i$}
\FAVert(10.,10.){1}

\end{feynartspicture}
\vspace{-1em}
\caption{Some generic \onel\ diagrams with counterterm insertion for the
  Higgs-boson self-energies ($\phi = h, H, A$),
$\;i,j,k = 1,2$). Similar diagrams arise for the tadpole contributions.}
\label{fig:FD1LCT}
%\end{center}
\end{figure}
%%%%%%%%%%%%%%%%%%%%%%%% F I G U R E %%%%%%%%%%%%%%%%%%%%%%%%%%%%%%%%%%%%%%%%%

The genuine \twol\ Feynman diagrams to be
evaluated for the Higgs boson self-energies, $\hSi_{hh,hH,HH}$, 
are shown in \reffi{fig:FD2L}.
The diagrams with subloop renormalization are depicted in
\reffi{fig:FD1LCT}.
The diagrams and the corresponding amplitudes have been generated with
the package \fa~\cite{feynarts,famssm}. The further evaluation has
been done using the program \tc~\cite{twocalc}.

%%%%%%%%%%%%%%%%%%%%%%%%%%%%%%%%%%%%%%%%%%%%%%%%%%%%%%%%%%%%%%%%%%%%%%%%%%%%%%%
%%%%%%%%%%%%%%%%%%%%%%%%%%%%%%%%%%%%%%%%%%%%%%%%%%%%%%%%%%%%%%%%%%%%%%%%%%%%%%%

\smallskip
We now turn to the renormalization in the scalar-quark sector,
entering via the diagrams depicted in \reffi{fig:FD1LCT}. 
Since the two-loop self-energies are evaluated at
\order{\al_{t,b}\als} it is sufficient to
determine the counterterms 
induced by the strong interaction only.
The tree-level relations and further details about our notation can be
found in \citere{mhiggsFDalbals}. In the following $\msqi$, $i = 1,2$
denotes the two mass eigenvalues of the stop ($q = t$) and sbottom 
($q = b$) sector. The corresponding unitary diagonalization matrix is
denoted as $U_{\sq_{ij}}$, $i,j = 1,2$.

%%%%%%%%%%%%%%%%%%%%%%%%%%%%%%%%%%%%%%%%%%%%%%%%%%%%%%%%%%%%%%%%%%%%%%%%%%%%%%%
%%%%%%%%%%%%%%%%%%%%%%%%%%%%%%%%%%%%%%%%%%%%%%%%%%%%%%%%%%%%%%%%%%%%%%%%%%%%%%%

\subsection{Renormalization of the top and scalar-top sector }
\label{subsec:stoprenorm}

The $t/\Stop$ sector contains four independent parameters:
the top-quark mass $\mt$, the stop masses $\mste$ and $\mstz$, and
either the squark mixing angle $\tst$ or, equivalently, the
trilinear coupling $\At$. Accordingly, the renormalization of this sector 
is performed by introducing four counterterms that are determined by
four independent renormalization conditions.
The following renormalization conditions are imposed (see also
\citere{hr}).  
\begin{itemize}
\item[(i)] 
On-shell renormalization of the top-quark mass
\item[(ii)] 
On-shell renormalization of the stop masses
\item[(iii)] 
The counterterm for the mixing angle, $\tst$:
$\de \tst = 
\KKL\re \Si_{\Stop_{12}}(m_{{\Stop}_{1}}^2)+\re 
\Si_{\Stop_{12}}(m_{{\Stop}_{2}}^2) \KKR / \KKL2(\mste^2-\mstz^2)\KKR \;$ .
\end{itemize}
Having already specified $\de \tst$,
the $A_t$ counterterm cannot be defined
independently but follows from the relation
$\sin 2 \tst = \KKL 2 \mt ( \At - \mu \cot \be) \KKR /
[ \mste^2 -\mstz^2 ]$ .

%%%%%%%%%%%%%%%%%%%%%%%%%%%%%%%%%%%%%%%%%%%%%%%%%%%%%%%%%%%%%%%%%%%%%%%%%%%%%%%
%%%%%%%%%%%%%%%%%%%%%%%%%%%%%%%%%%%%%%%%%%%%%%%%%%%%%%%%%%%%%%%%%%%%%%%%%%%%%%%

\subsection{Renormalization of the bottom and scalar-bottom sector}
\label{subsec:sbotrenorm}

Because of $SU(2)$-invariance 
the soft-breaking parameters for the left-handed stops and
sbottoms are identical, and thus
the $\Stop$~and $\Sbot$~masses are not
independent but connected via the relation
\begin{align}
 \cosQtb \msbe^2 + \sinQtb \msbz^2 = 
 \cosQtt \mste^2 + \sinQtt \mstz^2 +
\mb^2 - \mt^2 - M_W^2 \cos (2 \beta)~.
\label{MSb1gen}
\end{align}
Since the $\Stop$~masses have already been renormalized
on-shell, only one of the $\Sbot$~mass
counterterms can be determined independently. In the following, the
$\Sbotz$~mass is chosen as the pole mass, yielding
$\de \msbz^2 = \re \Si_{\Sbot_{22}}(\msbz^2)$, 
whereas the counterterm for $\msbe$ is determined 
as a combination of other counterterms, 
according to
\begin{align} 
\de \msbe^2 &= 
\frac{1}{\cos^2 \tsb} \Bigl(
  \cos^2 \tst \de \mste^2   
 + \sin^2 \tst \de \mstz^2 
 - \sin^2 \tsb \de \msbz^2
 - \sin 2 \tst (\mste^2 -\mstz^2)\de \tst \nonumber
 \\ %[1.5mm]
& \quad 
 + \sin 2 \tsb (\msbe^2 -\msbz^2)\de \tsb 
-  2 \mt\, \de\mt + 2 \mb\, \de \mb   \Bigr)~.
\label{ms1CT}
\end{align}
Consequently, the numerical value of $\msbe$ does not
correspond to the pole mass. The pole mass can be obtained 
from $\msbe$ via a finite shift of $\mathcal O(\als)$
(see e.g.~\citere{delrhosusy2loop}).

There are three more parameters with counterterms to be determined: 
the $b$-quark mass $\mb$, the mixing angle $\tsb$,
and the trilinear coupling~$\Ab$. They are connected via
\begin{align}
\label{mixingangleAparametermbrelation}
\sin 2 \tsb = \KKL 2 \mb ( \Ab - \mu\tb) \KKR / [ \msbe^2 -\msbz^2 ] \, ,
\end{align} 
which reads in terms of counterterms
\begin{align}
\label{deltaSbot}
2 \cosZtb\; \de\tsb &= \sinZtb \frac{\de\mb}{\mb} 
                   + \frac{2\mb\,\de\Ab}{\msbe^2 - \msbz^2}
                   - \sinZtb \frac{\de\msbe^2 - \de\msbz^2}
                                  {\msbe^2 - \msbz^2}~.
\end{align}
Only two of the three counterterms, $\de\mb$, $\de\tsb$, $\de\Ab$ can
be treated as independent, which offers a variety of choices.
In the following,
four different renormalization schemes, collected in \refta{tab:sbotren},
will be investigated.
Details about the four schemes can be found in \citere{mhiggsFDalbals}.

%%%%%%%%%%%%%%%%%%%%%%%%% T A B L E %%%%%%%%%%%%%%%%%%%%%%%%%%%%%%%%%%%%%%%%%%%
\begin{table}[!htb]
\BC
\caption {Summary of the four renormalization schemes for the bottom
quark/squark sector. Blank entries indicate
dependent quantities. For more details see \citere{mhiggsFDalbals}.} 
\begin{tabularx}{16.5cm}{|c||X|X|X|X|}
\hline 
scheme & $\msbz^2$ & $\mb$  & $\Ab$ & $\tsb$ \\ \hline\hline
analogous to $t/\Stop$ sector (``$\mb$ OS'') & on-shell & 
on-shell  & & on-shell\\\hline
\drbar\ bottom-quark mass  (``$\mb$ \drbar'') & on-shell &
\drbar   & \drbar  & \\\hline 
\drbar\ mixing angle and $\Ab$ (``$\Ab$, $\tsb$ \drbar'') & on-shell &   & 
\drbar  & \drbar  \\\hline 
on-shell mixing angle and $\Ab$ (``$\Ab$, $\tsb$ OS'') & on-shell &   &
on-shell & on-shell \\\hline 
\end{tabularx}
\label{tab:sbotren}
\EC
\end{table} 
%%%%%%%%%%%%%%%%%%%%%%%%% T A B L E %%%%%%%%%%%%%%%%%%%%%%%%%%%%%%%%%%%%%%%%%%%

%%%%%%%%%%%%%%%%%%%%%%%%%%%%%%%%%%%%%%%%%%%%%%%%%%%%%%%%%%%%%%%%%%%%%%%%%%%%%%%
%%%%%%%%%%%%%%%%%%%%%%%%%%%%%%%%%%%%%%%%%%%%%%%%%%%%%%%%%%%%%%%%%%%%%%%%%%%%%%%

\subsection{Resummation in the \boldmath{$b/\Sbot$} sector}
\label{subsec:botresum}

The relation between the bottom-quark mass and the Yukawa coupling $h_b$,
which in lowest order reads $\mb = h_b v_1/\sqrt{2}$, receives radiative
corrections proportional to $h_b v_2 = h_b \tb \, v_1$. Thus, large
$\tb$-enhanced contributions can occur, which need to be properly taken
into account. As shown in \citeres{deltamb1,deltamb} the leading terms of 
\order{\alb(\als\tb)^n} can be resummed by using an appropriate
effective bottom Yukawa coupling. 

Accordingly, an effective bottom-quark mass, 
$\mb^{\drbarm, \text{MSSM}}$, is obtained by extracting
its UV-finite $\tb$-enhanced contribution $\De \mb$ 
(which enters through $\Si_{b_S}$) and writing it
as $1/(1 + \De \mb)$ into the  
denominator. In this way the leading powers of $(\als\tb)^n$ are
correctly resummed~\cite{deltamb1,deltamb}. This yields
\begin{equation}
\label{eq:mbdrbarresum}
\mb^{\drbarm, \text{MSSM}}(\mu^{\drbarm}) = 
  [ \mb^{\msbarm}(M_Z) b^{\text{shift}}  + \edz  \mb
  (\Si^{\rm fin}_{b_L}({\mb}^2) + \Si^{\rm fin}_{{b}_R} ({\mb}^2))
+ \mb\, \widetilde{\Si}^{\rm fin}_{b_S}(\mb^2) ] / 
[ 1 + \De \mb ]~,
\end{equation}
where $\widetilde{\Si}_{b_S} \equiv \Si_{b_S} + \De \mb$ denotes the 
non-enhanced remainder of the scalar $b$-quark self-energy at
\order{\als}, and $b^{\text{shift}}$ is given in by
$
b^{\text{shift}} \equiv \Bigl[1 +
\frac{\alpha_s}{\pi} \Bigl(\frac{4}{3} - \ln
\frac{(\mb^{\msbarm})^2}{M_Z^2} \Bigr)\Bigr]
$.
The $\tb$-enhanced scalar part of the
$b$-quark self-energy, $\De \mb$, is given at \order{\als} by
\begin{align}
\label{eq:deltamb}
\De \mb = - 2/(3 \pi) \als\tb\, \mu\, \mgl\,
 \frac{\msbe^2 \msbz^2 \log(\msbz^2/\msbe^2) +
        \msbe^2 \mgl^2  \log(\msbe^2/\mgl^2) +
        \mgl^2 \msbz^2  \log(\mgl^2/\msbz^2)}
       {(\msbe^2 - \mgl^2) (\mgl^2 - \msbz^2) (\msbz^2 - \msbe^2)}~.
\end{align}

We incorporate the effective bottom-quark mass of \refeq{eq:mbdrbarresum}
into our one-loop results for the 
renormalized Higgs boson self-energies.
In this way the leading effects
of \order{\alb\als} are absorbed into the one-loop result. 
We refer to the genuine two-loop contributions,
which go beyond this improved one-loop result, as ``subleading
\order{\alb\als} corrections'' in the following.

%%%%%%%%%%%%%%%%%%%%%%%%%%%%%%%%%%%%%%%%%%%%%%%%%%%%%%%%%%%%%%%%%%%%%%%%%%%%%%%
%%%%%%%%%%%%%%%%%%%%%%%%%%%%%%%%%%%%%%%%%%%%%%%%%%%%%%%%%%%%%%%%%%%%%%%%%%%%%%%

\section{NUMERICAL RESULTS}
\label{sec:numres}

For our numerical analysis we use the following parameters (if not
indicated differently): 
$\mt = 174.3 \gev$, $\mb^{\msbarm} = 2.94 \gev$, $\MA = 700 \gev$, 
$\mu = -1000$, $\tb = 50$, $\msusy = 1000 \gev$, $A_f = 2 \msusy$,
$M_2 = 2 M_1 = 100 \gev$, $\mgl = 1000 \gev$, $\mu^{\drbarm} = \mt$.
Large values of $\tb$ and $|\mu|$ are chosen in order to
illustrate possibly large effects in the $b/\Sbot$ sector.
The inclusion of all known corrections and the current experimental 
central top quark mass value of $\mt = 178.0 \gev$ in our analysis
would yield an 
increase in $\Mh$ of \order{8 \gev}~\cite{mhiggsAEC}.
Therefore the mass values given
in our numerical analysis should not be viewed as predictions of
$\Mh$; they are rather illustrations of the $\als$-corrections
to the bottom Yukawa contributions at the two-loop level. 

\definecolor{lightblue}{cmyk}{1,0,0,0}
\definecolor{Blue}{rgb}{0,0,1}
\definecolor{Red}{named}{Red}
\definecolor{Green}{rgb}{0,0.9,0.2}
\definecolor{Black}{named}{Black}
\definecolor{Magenta}{named}{Magenta}
\definecolor{Royal}{named}{RoyalBlue}
\definecolor{Orange}{named}{Orange}
\definecolor{Purple}{named}{Purple}
\definecolor{Mahogany}{named}{Mahogany}
\definecolor{Brown}{named}{Brown}

\psfrag{MHH [GeV]}{{$\MH$ [GeV]}}
\psfrag{Mh0 [GeV]}{{$\Mh$ [GeV]}}
\psfrag{Delta Mh0 [GeV]}{{$\De\Mh$ [GeV]}}
\psfrag{MA0 [GeV]}{ $\MA$ [GeV]}
\psfrag{tan beta}{\raisebox{0.ex}{{$\tb$}}}
\psfrag{MUE [GeV]}{$\mu$ [GeV]}
\psfrag{MGl [GeV]}{$\mgl$ [GeV]}
\psfrag{MUE = -1000 GeV}{$\mu = -1000$ GeV}
\psfrag{MUE = 1000 GeV}{$\mu = 1000$ GeV}
\psfrag{MA0 = 120 GeV}{$\MA = 120$ GeV} 
\psfrag{MA0 = 700 GeV}{$\MA = 700$ GeV} 
\psfrag{MGl = 1000 GeV}{$\mgl =  1000$ GeV}
\psfrag{MGl = 1500 GeV}{$\mgl =  1500$ GeV}
\psfrag{MUE = -1000 GeV, MA0 = 120 GeV, MGl = 1000 GeV}
       {$\mu = -1000$ GeV, $\MA = 120$ GeV, $\mgl = 1000$ GeV}
\psfrag{MUE = 1000 GeV, MA0 = 120 GeV, MGl = 1000 GeV}
       {$\mu = 1000$ GeV, $\MA = 120$ GeV, $\mgl = 1000$ GeV}  
\psfrag{MUE = 1000 GeV, MA0 = 700 GeV, MGl = 1000 GeV}
       {$\mu = 1000$ GeV, $\MA = 700$ GeV, $\mgl = 1000$ GeV}  
\psfrag{MUE = -1000 GeV, TB = 50, MGl = 1000 GeV}
       {$\mu = -1000$ GeV, $\tb = 50$, $\mgl = 1000$ GeV}  
\psfrag{TB = 50}{ $\tan \beta = 50$}
\psfrag{TB = 50 GeV}{ $\tan \beta = 50$}
\psfrag{TB = 50, MA0 = 120 GeV, MGl = 1000 GeV}
       {$\tb = 50$, $\MA = 120$ GeV, $\mgl = 1000$ GeV}  
\psfrag{TB = 50, MA0 = 700 GeV, MGl = 1000 GeV}
       {$\tb = 50$, $\MA = 700$ GeV, $\mgl = 1000$ GeV}  

\psfrag{O(a_s a_t) with 1234567}
       {\order{\alt\als} with $\mb^{\drbarm,\text{MSSM}}$}
\psfrag{scheme1}
       {\hspace*{-3.4cm}\color{lightblue}{\order{\alb\als} : scheme: $\mb$ OS}}
\psfrag{scheme2}
       {\hspace*{-3.4cm}\Red{\order{\alb\als} : scheme: $\mb\,\drbarm$}}
\psfrag{scheme3}
       {\hspace*{-3.4cm}\Blue{\order{\alb\als} : scheme: $\Ab,\;\tsb \drbarm$}}
\psfrag{scheme4}
       {\hspace*{-3.4cm}\Green{\order{\alb\als} : scheme: $\Ab,\;\tsb$ OS}}

\psfrag{scheme1b}
       {\hspace*{-0.0cm}{\order{\alt\als} ($\mb^{\drbarm,\text{MSSM}}$)}}
\psfrag{scheme2b}
       {\hspace*{-0.0cm}\Red{\order{\alb\als} : $\mb\,\drbarm$}}
\psfrag{scheme3b}
       {\hspace*{-0.0cm}\Blue{\order{\alb\als} : $\Ab,\;\tsb~\drbarm$}}
\psfrag{scheme4b}
       {\hspace*{-0.0cm}\Green{\order{\alb\als} : $\Ab,\;\tsb$ OS}}
\psfrag{scheme2b1L}
       {\hspace*{-0.0cm}{\order{\alt\als} : $\mb\,\drbarm$ for \order{\alb}}}
\psfrag{scheme3b1L}
       {\hspace*{-0.0cm}{\order{\alt\als} : $\Ab,\;\tsb~\drbarm$ for \order{\alb}}}
\psfrag{scheme4b1L}
       {\hspace*{-0.0cm}{\order{\alt\als} : $\Ab,\;\tsb$ OS for \order{\alb}}}

\psfrag{O(a_s a_t) default param.}{}
\psfrag{UIF-type scheme O(a_s a_t)}
       {\hspace*{-0.0cm}\order{\alt\als} : $\Ab,\;\tsb$ OS for \order{\alb}}
\psfrag{UIF-type scheme} 
       {\hspace*{-0.3cm}\order{\alb\als} : $\Ab,\;\tsb$ OS}
\psfrag{with Pietros code O(a_s a_t)}
      {\hspace*{-0.4cm}\order{\alt\als} : $\msb,\;\Ab$~OS for \order{\alb}}
\psfrag{with Pietros code}  
       {\hspace*{-0.7cm}\order{\alb\als} : $\msb,\;\Ab$~OS}

\psfrag{diffscheme2mehrplatz}
       {\hspace*{+1.0cm}{\order{\alb\als} : $\mb\,\drbarm$}}
\psfrag{diffscheme3}
       {\hspace*{-1.5cm}{\order{\alb\als} : $\Ab,\;\tsb~\drbarm$}}
\psfrag{diffscheme4}
       {\hspace*{-1.5cm}{\order{\alb\als} : $\Ab,\;\tsb$ OS}}

In \citere{mhiggsFDalbals} it has been shown that the ``$\mb$~OS''
scheme should be discarded.
The reason for the problematic behavior of this scheme is
easy to understand. The renormalization condition in the ``$\mb$ OS''
scheme is a condition on the sbottom mixing angle $\tsb$ and thus on the 
combination $(\Ab - \mu \tb)$.
In parameter regions where
$\mu \tb$ is much larger than $\Ab$, the counterterm $\de\Ab$ receives
a very large finite shift when calculated from the counterterm 
$\de\tsb$. This problem is
avoided in the other renormalization schemes introduced in 
\refta{tab:sbotren}, where the renormalization condition is applied
directly to $\Ab$, rather than deriving $\de\Ab$ from the
renormalization of the mixing angle.

%%%%%%%%%%%%%%%%%%%%%%%% F I G U R E %%%%%%%%%%%%%%%%%%%%%%%%%%%%%%%%%%%%%%%%%
\begin{figure}[htb!]
\begin{center}
\epsfig{figure=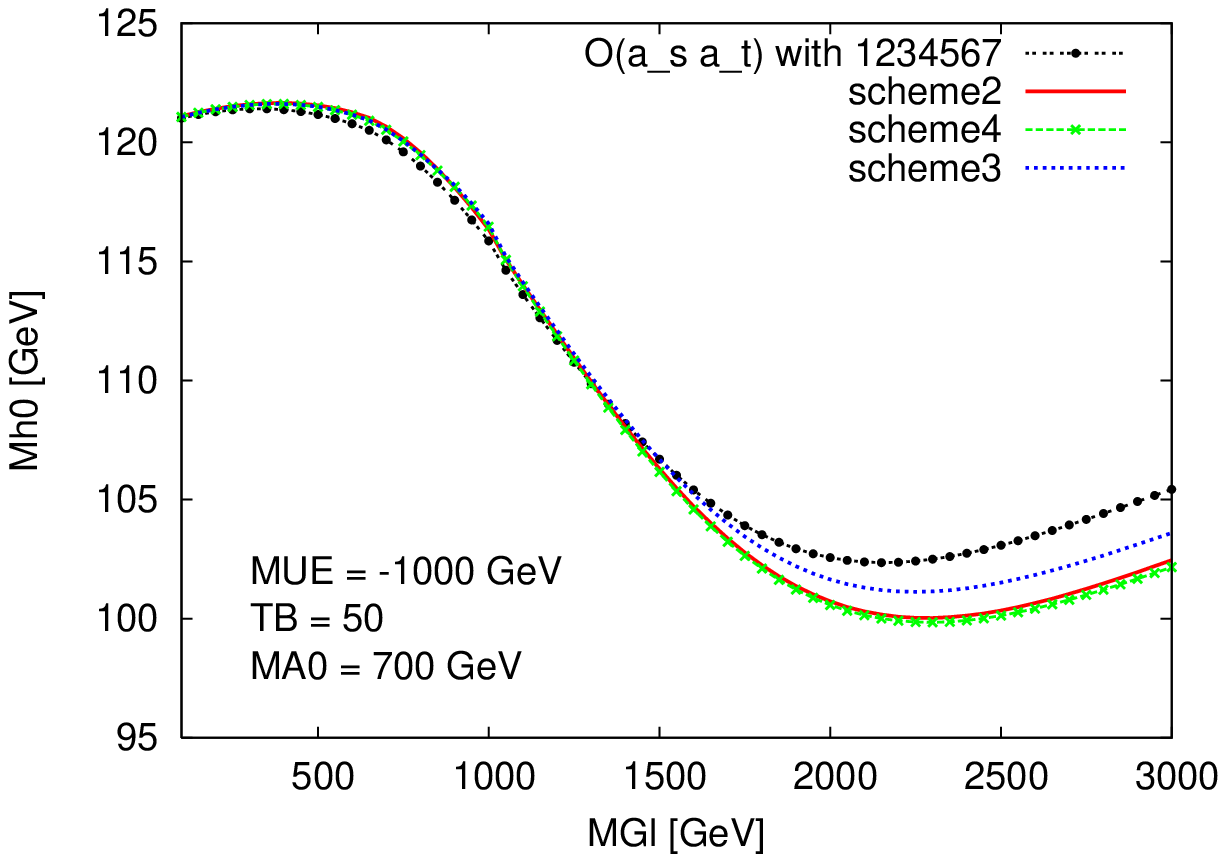, width=13cm,height=7.5cm}
\caption{
$\Mh$ as a function of $\mgl$.
}
\label{fig:mhmgl}
\end{center}
\end{figure}
%%%%%%%%%%%%%%%%%%%%%%%% F I G U R E %%%%%%%%%%%%%%%%%%%%%%%%%%%%%%%%%%%%%%%%%

In \reffi{fig:mhmgl} it can be seen that the behavior of the
subleading corrections in the three remaining renormalization schemes
strongly depends on the choice of $\mgl$. 
For $\mgl \lsim 1000 \gev$ all
schemes lead to an increase of $\Mh$ from the subleading
\order{\alb\als} corrections. For $\mgl \gsim 1500 \gev$, on the other
hand, all schemes lead to a decrease, where the size of the individual
corrections also strongly varies with $\mgl$. Accordingly, the relative
size of the corrections in the different schemes also
varies with $\mgl$. Corrections up to about $3 \gev$ are possible. 
The differences between the three schemes
are of \order{2 \gev} for large $\mgl$. 
It should be noted that the effects of the higher-order corrections to
$\Mh$ do not decouple with large $\mgl$. The corrections at
\order{\alt\als}~\cite{mhiggslong} as well as \order{\alb\als} grow
logarithmically in the renormalization schemes that we have adopted. 

%%%%%%%%%%%%%%%%%%%%%%%% F I G U R E %%%%%%%%%%%%%%%%%%%%%%%%%%%%%%%%%%%%%%%%%
\begin{figure}[hb!]
\begin{center}
\epsfig{figure=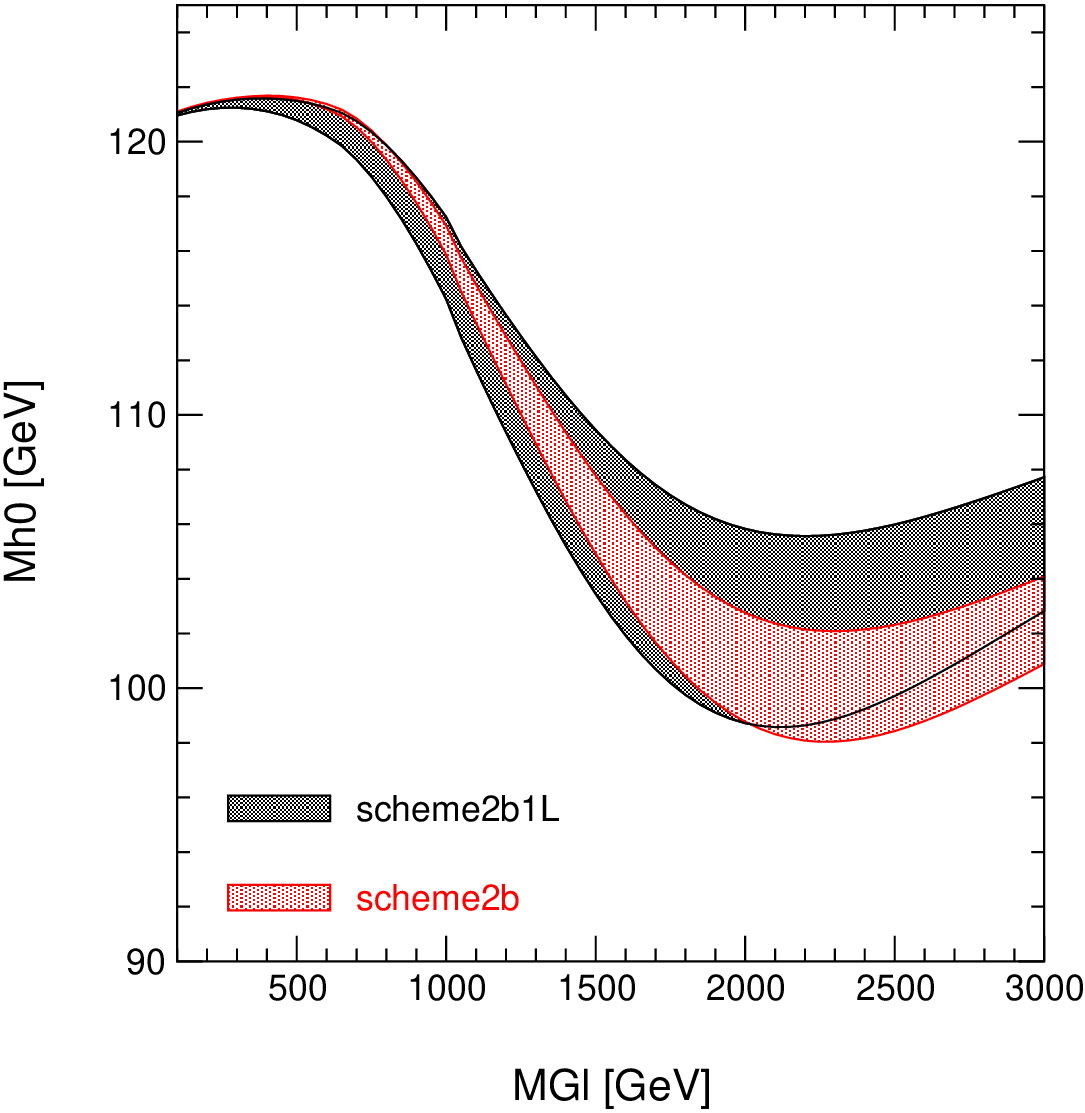, width=11cm,height=6.5cm}
\caption{
$\mudim$ dependence of $\Mh$ as a function of $\mgl$. The black area
corresponds to the \order{\alt\als} result including resummation,
i.e.\ the result without the subleading two-loop \order{\alb\als} terms.
The parameters are the same as in \reffi{fig:mhmgl}.
}
\label{fig:mhmudim}
\end{center}
\end{figure}
%%%%%%%%%%%%%%%%%%%%%%%% F I G U R E %%%%%%%%%%%%%%%%%%%%%%%%%%%%%%%%%%%%%%%%%

The $\mudim$ variation is shown in \reffi{fig:mhmudim}. The leading
contribution (the \order{\alt\als} 
result including resummation) is shown as the
dark shaded (black) band. The results including the subleading
corrections in the ``$\mb$ \drbar'' scheme are shown as a light shaded
(red) band. It can be seen that the variation with $\mudim$ is
strongly reduced by the inclusion of the subleading contributions. 
The variation with $\mudim$ within the ``$\mb$ \drbar'' scheme is tiny
for $\mgl \lsim 500 \gev$, and reaches $\pm 2 \gev$  for large
$\mgl$ values. 
Thus, the $\mudim$ variation causes a similar shift in $\Mh$ as the
comparison between the three renormalization schemes discussed above.

The comparison of the results in the different schemes that we have
analyzed and the investigation of the renormalization scale dependence 
give an indication of the possible size of missing higher-order
corrections in the $b/\Sbot$~sector. For $\mu > 0$ (see
\citere{mhiggsFDalbals}) the higher-order
corrections from the $b/\Sbot$~sector (beyond \order{\alb\als}) appear
to be sufficiently well under control.
For $\mu < 0$, on the other hand, sizable
higher-order corrections from the $b/\Sbot$~sector are possible. The
size of the individual corrections and also the difference between the 
analyzed schemes varies significantly with the relevant parameters,
$\mu$, $\tb$, $\mgl$ and $\MA$. We estimate the uncertainty from missing
higher-order corrections in the $b/\Sbot$~sector to be about $2 \gev$ in 
this region of parameter space.

The results obtained will be implemented into the Fortran code
\fh~\cite{mhiggslong,mhiggsAEC,feynhiggs}.

%%%%%%%%%%%%%%%%%%%%%%%%%%%%%%%%%%%%%%%%%%%%%%%%%%%%%%%%%%%%%%%%%%%%%%%%%%%%%%%
%%%%%%%%%%%%%%%%%%%%%%%%%%%%%%%%%%%%%%%%%%%%%%%%%%%%%%%%%%%%%%%%%%%%%%%%%%%%%%%

%%%%%%%%%%%%%%%%%%%%%%%%%%%%%%%%%%%%%%%%%%%%%%%%%%%%%%%%%%%%%%%%%%%%%%%%%%%%%%%
%%%%%%%%%%%%%%%%%%%%%%%%%%%%%%%%%%%%%%%%%%%%%%%%%%%%%%%%%%%%%%%%%%%%%%%%%%%%%%%

\end{document}

%% file: paperdef.tex
\makeatletter
\def\citer{\@ifnextchar [{\@tempswatrue\@citexr}{\@tempswafalse\@citexr[]}}
 
\def\@citexr[#1]#2{\if@filesw\immediate\write\@auxout{\string\citation{#2}}\fi
  \def\@citea{}\@cite{\@for\@citeb:=#2\do
    {\@citea\def\@citea{--\penalty\@m}\@ifundefined
       {b@\@citeb}{{\bf ?}\@warning
       {Citation `\@citeb' on page \thepage \space undefined}}%
\hbox{\csname b@\@citeb\endcsname}}}{#1}}
\makeatother

\def\refeq#1{\mbox{(\ref{#1})}}

\def\reffi#1{\mbox{Fig.~\ref{#1}}}

\def\refta#1{\mbox{Tab.~\ref{#1}}}
\def\citere#1{\mbox{Ref.~\cite{#1}}}
\def\citeres#1{\mbox{Refs.~\cite{#1}}}

\def\xtilde#1{%
  \setbox0\hbox{$\tilde#1$}%
  \rlap{\raise\ht0\hbox{\tiny$_{\,(\;\,)}$}}%
  \tilde#1%
}

\newcommand{\mste}{m_{\tilde{t}_1}}
\newcommand{\mstz}{m_{\tilde{t}_2}}

\newcommand{\msb}{m_{\tilde{b}}}

\newcommand{\msbe}{m_{\tilde{b}_1}}
\newcommand{\msbz}{m_{\tilde{b}_2}}

\newcommand{\At}{A_t}
\newcommand{\Ab}{A_b}

\newcommand{\msusy}{M_{\text{SUSY}}}

\newcommand{\msqi}{m_{\tilde{q}_i}}

\newcommand{\mhtree}{m_{h,{\rm tree}}}
\newcommand{\mHtree}{m_{H,{\rm tree}}}

\newcommand{\mudim}{\mu^{\drbarm}}

\newcommand{\msbarm}{\overline{\rm{MS}}}
\newcommand{\drbar}{$\overline{\rm{DR}}$}
\newcommand{\drbarm}{\overline{\rm{DR}}}

\def\order#1{${\cal O}(#1)$}
\newcommand{\cp}{{\cal CP}}

\newcommand{\edz}{\frac{1}{2}}

\newcommand{\twol}{two-loop}
\newcommand{\onel}{one-loop}

\newcommand{\tc}{{\em TwoCalc}}
\newcommand{\fa}{{\em FeynArts}}
\newcommand{\fh}{{\em FeynHiggs}}

\newcommand{\MA}{M_A}
\newcommand{\mh}{m_h}
\newcommand{\mH}{m_H}
\newcommand{\Mh}{M_h}
\newcommand{\MH}{M_H}

\newcommand{\mt}{m_{t}}

\newcommand{\mb}{m_{b}}

\newcommand{\mgl}{m_{\tilde{g}}}
\newcommand{\sq}{\tilde{q}}

\newcommand{\Stop}{\tilde{t}}

\newcommand{\Sbot}{\tilde{b}}

\newcommand{\Sbotz}{\tilde{b}_2}

\newcommand{\tst}{\theta_{\tilde{t}}}
\newcommand{\tsb}{\theta_{\tilde{b}}}

\newcommand{\tsf}{\theta\kern-.20em_{\tilde{f}}}
\newcommand{\tsfp}{\theta\kern-.20em_{\tilde{f}\prime}}
\newcommand{\tsq}{\theta\kern-.15em_{\tilde{q}}}

\newcommand{\sinQtt}{\sin^2\tst}

\newcommand{\sinQtb}{\sin^2\tsb}
\newcommand{\sinZtb}{\sin 2\tsb}

\newcommand{\cosQtt}{\cos^2\tst}

\newcommand{\cosQtb}{\cos^2\tsb}
\newcommand{\cosZtb}{\cos 2\tsb}

\newcommand{\KKL}{\left[}
\newcommand{\KKR}{\right]}

\newcommand{\VL}{\left( \begin{array}{c}}
\newcommand{\VR}{\end{array} \right)}
\newcommand{\ML}{\left( \begin{array}{cc}}
\newcommand{\MLd}{\left( \begin{array}{ccc}}
\newcommand{\MLv}{\left( \begin{array}{cccc}}
\newcommand{\MR}{\end{array} \right)}

\newcommand{\re}{\mathop{\rm Re}}

\newcommand{\tb}{\tan \beta}

\newcommand{\gev}{\,\, {\rm GeV}}

\newcommand{\BC}{\begin{center}}
\newcommand{\EC}{\end{center}}
\newcommand{\BE}{\begin{equation}}
\newcommand{\EE}{\end{equation}}
\newcommand{\BEA}{\begin{eqnarray}}
\newcommand{\BEAnn}{\begin{eqnarray*}}
\newcommand{\EEA}{\end{eqnarray}}
\newcommand{\EEAnn}{\end{eqnarray*}}

\newcommand{\id}{{\rm 1\kern-.12em
\rule{0.3pt}{1.5ex}\raisebox{0.0ex}{\rule{0.1em}{0.3pt}}}}
\newcommand{\lsim}
{\;\raisebox{-.3em}{$\stackrel{\displaystyle <}{\sim}$}\;}
\newcommand{\gsim}
{\;\raisebox{-.3em}{$\stackrel{\displaystyle >}{\sim}$}\;}

\def\al{\alpha}

\def\als{\alpha_s}
\def\alt{\alpha_t}
\def\alb{\alpha_b}

\def\be{\beta}

\def\de{\delta}

\def\De{\Delta}

\def\Si{\Sigma}

\def\hSi{\hat{\Sigma}}

\def\3{\ss}

%% file: lcwsSLACHiggs.bbl
\begin{thebibliography}{99}

\bibitem{susy} H.P.~Nilles, 
               {\em Phys.\ Rep.} {\bf 110} (1984) 1; 
               %%CITATION = PRPLC,110,1;%%
               H.E.~Haber and G.L.~Kane, 
               {\em Phys.\ Rep.} {\bf 117} (1985) 75; 
               %%CITATION = PRPLC,117,75;%%
               R.~Barbieri, 
               {\em Riv.\ Nuovo Cim.} {\bf 11} (1988) 1. 
               %%CITATION = RNCIB,11,1;%%

\bibitem{LEPHiggsSM} [LEP Higgs working group], 
                     {\em Phys. Lett.} {\bf B 565} (2003) 61, 
                     hep-ex/0306033.
                     %%CITATION = HEP-EX 0306033;%%

\bibitem{LEPHiggsMSSM} [LEP Higgs working group],
                    hep-ex/0107030;
                    %%CITATION = HEP-EX 0107030;%%
                    hep-ex/0107031;
                    %%CITATION = HEP-EX 0107031;%%
                    LHWG-Note 2004-01,
                    see:\\ {\tt lephiggs.web.cern.ch/LEPHIGGS/papers/} .

\bibitem{LHCHiggs} ATLAS Collaboration, 
  {\em Detector and Physics Performance Technical Design Report},
  CERN/LHCC/99-15 (1999), see:
 {\tt atlasinfo.cern.ch/Atlas/GROUPS/PHYSICS/TDR/access.html}~;\\
 CMS Collaboration, see:
 {\tt cmsinfo.cern.ch/Welcome.html/CMSdocuments/CMSplots/}~.

\bibitem{HcoupLHCSM} M.~D\"uhrssen, S.~Heinemeyer, H.~Logan, D.~Rainwater, 
                     G.~Weiglein and D.~Zeppenfeld,
                     {\em Phys. Rev.} {\bf D 70} (2004) 113009,
                     hep-ph/0406323.
                     %%CITATION = HEP-PH 0406323;%%

\bibitem{tesla} J.~Aguilar-Saavedra et al.,
                TESLA TDR Part~3, 
%                ``Physics at an $e^+e^-$ Linear Collider'', 
                hep-ph/0106315,
                %%CITATION = HEP-PH 0106315;%%
                see: {\tt tesla.desy.de/tdr/} .
             
\bibitem{orangebook} T.~Abe et al.
                     [American Linear Collider Working Group Collaboration],
%                     {\it Resource book for Snowmass 2001}, 
                     hep-ex/0106056.
                     %%CITATION = HEP-EX 0106056;%%

\bibitem{acfarep} K.~Abe et al. 
                  [ACFA Linear Collider Working Group Collaboration],
                  hep-ph/0109166.
                  %%CITATION = HEP-PH 0109166;%%

\bibitem{tbexcl} S.~Heinemeyer, W.~Hollik and G.~Weiglein, 
                 {\em JHEP} {\bf 0006} (2000) 009,
                 hep-ph/9909540.
                 %%CITATION = HEP-PH 9909540;%%

\bibitem{PomssmRep} S.~Heinemeyer, W.~Hollik and G.~Weiglein, 
                    hep-ph/0412214.
                    %%CITATION = HEP-PH 0412214;%%

\bibitem{mhiggsAEC} G.~Degrassi, S.~Heinemeyer, W.~Hollik,
                    P.~Slavich and G.~Weiglein,
                    {\em Eur. Phys. J.} {\bf C 28} (2003) 133,
                    hep-ph/0212020.
                    %%CITATION = HEP-PH 0212020;%%

\bibitem{habilSH} S.~Heinemeyer, 
                  hep-ph/0407244.
                  %%CITATION = HEP-PH 0407244;%%

\bibitem{mhiggsAWB} A.~Djouadi,
                    hep-ph/0503173.
                    %%CITATION = HEP-PH 0503173;%%

\bibitem{mhiggsEP4} A.~Brignole, G.~Degrassi, P.~Slavich and F.~Zwirner,
                    {\em Nucl. Phys.} {\bf B 643} (2002) 79,
                    hep-ph/0206101.
                    %%CITATION = HEP-PH 0206101;%%

\bibitem{mhiggsFDalbals} S.~Heinemeyer, W.~Hollik, H.~Rzehak  and G.~Weiglein,
                         to appear in {\em Eur. Phys. J.} {\bf C}, 
                         hep-ph/0411114.
                         %%CITATION = HEP-PH 0411114;%%

\bibitem{deltamb1} T.~Banks, 
                   {\em Nucl.\ Phys.} {\bf B 303} (1988) 172;
                   %%CITATION = NUPHA,B303,172;%%
                   L.~Hall, R.~Rattazzi and U.~Sarid,
                   {\em Phys.\ Rev.} {\bf D 50} (1994) 7048, 
                   hep-ph/9306309;
                   %%CITATION = HEP-PH 9306309;%%
                   R.~Hempfling, 
                   {\em Phys.\ Rev.} {\bf D 49} (1994) 6168;
                   %%CITATION = PHRVA,D49,6168;%%
                   M.~Carena, M.~Olechowski, S.~Pokorski and C.~Wagner,
                   {\em Nucl.\ Phys.}\ {\bf B 426} (1994) 269, 
                   hep-ph/9402253.
                   %%CITATION = HEP-PH 9402253;%%

\bibitem{deltamb} M.~Carena, D.~Garcia, U.~Nierste and C.~Wagner,
                  {\em Nucl. Phys.} {\bf B 577} (2000) 577,
                  hep-ph/9912516;
                  %%CITATION = HEP-PH 9912516;%%
                  H.~Eberl, K.~Hidaka, S.~Kraml, W.~Majerotto and
                  Y.~Yamada,
                  {\em Phys. Rev.} {\bf D 62} (2000) 055006,
                  hep-ph/9912463.
                  %%CITATION = HEP-PH 9912463;%%

\bibitem{effpotfull} S.~Martin,
                     {\em Phys. Rev.} {\bf D 65} (2002) 116003,
                     hep-ph/0111209;
                     %%CITATION = HEP-PH 0111209;%%
                     {\em Phys. Rev.} {\bf D 66} (2002) 096001,
                     hep-ph/0206136;
                     %%CITATION = HEP-PH 0206136;%%
                     Phys. Rev. {\bf D 67} (2003) 095012, 
                     hep-ph/0211366;
                     %%CITATION = HEP-PH 0211366;%%
                     {\em Phys. Rev.} {\bf D 68} 075002 (2003), 
                     hep-ph/0307101; 
                     %%CITATION = HEP-PH 0307101;%%
                     {\em Phys. Rev.} {\bf D 70} (2004) 016005, 
                     hep-ph/0312092;
                     %%CITATION = HEP-PH 0312092;%%
                     {\em Phys. Rev.} {\bf D 71} (2005) 016012,
                     hep-ph/0405022.
                     %%CITATION = HEP-PH 0405022;%%

\bibitem{hhg} J.~Gunion, H.~Haber, G.~Kane and S.~Dawson,
              {\em The Higgs Hunter's Guide}, Addison-Wesley, 1990.

\bibitem{mhiggslong} S.~Heinemeyer, W.~Hollik and G.~Weiglein, 
                     {\em Eur. Phys. J.} {\bf C 9} (1999) 343, 
                     hep-ph/9812472.
                     %%CITATION = HEP-PH 9812472;%%

\bibitem{feynarts} J.~K\"ublbeck, M.~B\"ohm and A.~Denner, 
                   {\em Comp. Phys. Comm.} {\bf 60} (1990) 165;
                   %%CITATION = CPHCB,60,165;%%
                   T.~Hahn,
                   {\em Comput. Phys. Comm.} {\bf 140} (2001) 418,
                   hep-ph/0012260;
                   %%CITATION = HEP-PH 0012260;%%
                   The program is available via {\tt www.feynarts.de} .

\bibitem{famssm} T.~Hahn and C.~Schappacher,
                 {\em Comput. Phys. Comm.} {\bf 143} (2002) 54,
                 hep-ph/0105349.
                 %%CITATION = HEP-PH 0105349;%%

\bibitem{twocalc} G.~Weiglein, R.~Scharf and M.~B\"ohm,
                  {\em Nucl. Phys.} {\bf B 416} (1994) 606,
                  hep-ph/9310358;
                  %%CITATION = HEP-PH 9310358;%%
G.~Weiglein, R.~Mertig, R.~Scharf and M.~B\"ohm, 
in {\it New Computing Techniques in Physics Research 2},
ed.~D.~Perret-Gallix (World Scientific, Singapore, 1992), p.~617.

\bibitem{hr} W. Hollik and H. Rzehak, 
             {\em Eur. Phys. J.} {\bf C 32} (2003) 127, 
             hep-ph/0305328.
             %%CITATION = HEP-PH 0305328;%%

\bibitem{delrhosusy2loop} A.~Djouadi, P.~Gambino, S.~Heinemeyer, W.~Hollik,
                          C.~J\"unger and G.~Weiglein,
                          {\em Phys. Rev. Lett.} {\bf 78} (1997) 3626,
                          hep-ph/9612363;
                          %%CITATION = HEP-PH 9612363;%%
                          {\em Phys. Rev.} {\bf D 57} (1998) 4179,
                          hep-ph/9710438.
                          %%CITATION = HEP-PH 9710438;%%

\bibitem{feynhiggs} S.~Heinemeyer, W.~Hollik and G.~Weiglein, 
                    {\em Comp. Phys. Comm.} {\bf 124} (2000) 76,
                    hep-ph/9812320; 
                    %%CITATION = HEP-PH 9812320;%%
                    hep-ph/0002213;
                    %%CITATION = HEP-PH 0002213;%% 
                 M.~Frank,T.~Hahn, S.~Heinemeyer, W.~Hollik and G.~Weiglein, 
                 {\em in preparation}.
                 The codes are accessible via
                 {\tt www.feynhiggs.de} .


\end{thebibliography}
